\documentclass[10pt, conference]{IEEEtran}

\usepackage{flushend}

\usepackage{enumitem}

\usepackage{booktabs}
\ifCLASSOPTIONcompsoc
  \usepackage[nocompress]{cite}
\else
  \usepackage{cite}
\fi
\ifCLASSINFOpdf
   \usepackage[pdftex]{graphicx}
  \DeclareGraphicsExtensions{.pdf,.jpeg,.png}
\else
\fi
\usepackage[cmex10]{amsmath}
\usepackage{algorithmic}
\ifCLASSOPTIONcompsoc
  \usepackage[caption=false,font=footnotesize,labelfont=sf,textfont=sf]{subfig}
\else
  \usepackage[caption=false,font=footnotesize]{subfig}
\fi
\usepackage{url}

\begin{document}
%
\title{Dynamic Multiparty Authentication of Data Analytics Services within Cloud Environments}
\author{\IEEEauthorblockN{Hussain Al-Aqrabi and Richard Hill}
\IEEEauthorblockA{Centre for Industrial Analytics\\
University of Huddersfield\\
Huddersfield, HD1 3DH, UK\\
Email: \{h.al-aqrabi, r.hill\}@hud.ac.uk}
}


%


\maketitle

\begin{abstract}
Approaches to  the provision of data analytics for businesses offer methods to analyse and model data, enabling informed decision making to improve business performance and profitability. Typically, analytics processing is an intensive task and the demand for business insight, on-demand, means that organisations make use of elastic cloud provisioned resources to host such services. However, within the shared domains of multi-tenant cloud computing, business data and models are exposed to greater security threats and compromised privacy, since an unauthorised user may be able to gain access to highly sensitive, consolidated business-critical information. Business analytics processes are often composed from orchestrated, collaborating services, which are consumed by users from multiple cloud systems (in different security realms), which need to be engaged dynamically at runtime. If heterogeneous cloud systems located in different security realms do not have direct authentication relationships, then it is a considerable technical challenge to enable secure collaboration. In order to address this security challenge, a new authentication framework is required to establish trust amongst business analytics service instances and users by distributing a common session secret to all participants of a session. We address this challenge by designing and implementing a secure multiparty authentication framework for dynamic interaction, for the scenario where members of different security realms express a need to access orchestrated services. This novel framework exploits the relationship of trust between session members in different security realms, to enable a user to obtain security credentials that access cloud resources in a remote realm. The mechanism assists cloud session users to authenticate their session membership, thereby improving the performance of authentication processes within multiparty sessions. We see applicability of this framework beyond multiple cloud infrastructure, to that of any scenario where multiple security realms has the potential to exist, such as the emerging Internet of Things (IoT).
\end{abstract}

\providecommand{\keywords}[1]
{
  \small	
  \textbf{\textit{Keywords---}} #1
}

\keywords{Cloud computing, analytics, security, multiparty interactions, Internet of Things}

%
\IEEEpeerreviewmaketitle

\section{Introduction}
It is common for modern organisations to utilise and benefit from the adoption of cloud and internet based computing. Users of cloud computing infrastructure can concentrate on the delivery of high value/profit services at high levels of Quality of Service (QoS), without being pre-occupied with large-scale investments in hardware and specialist skills for the support and maintenance of such systems\cite{albeshri2010,calheiros2011}. In particular the regular upgrades and expense of licenses of application software used to run business processes, related transactions and decision-support systems\cite{bieter2010} are a significant demand upon an organisation's resources.

Cloud-based systems are a fundamental enabling technology in this regard, that provide utility and ease with regard to universal availability and timely access. However, cloud computing has its own constraints and security considerations that need to be taken into account with regard to its effective deployment\cite{alaqrabi2014a}. 

Security is of paramount importance to a business organisation that consumes and generates data, and the awareness of the risks associated with data breaches means that ignorance of the approaches to managing cloud security is a significant inhibiting factor in the adoption of cloud environments\cite{sharma2018,chen2012}.

With the growing popularity of services delivered by cloud computing, it is important that both cloud providers and cloud users have the appropriate safeguards in place to ensure satisfactory security and protection of privacy\cite{choo2017,liu2015}. A number of researchers have contributed to strengthen security and privacy protection in cloud applications, and there are various cryptographic algorithms to address potential security and privacy problems in cloud\cite{ateniese2000,katz2003,rahulamathavan2014,yoon}. It follows that research topics related to cloud security have attracted tremendous research interest\cite{thilakanathan,song2016,arya,celesti2010b}.
\subsection{Business Intelligence and data analytics}
Business Intelligence (BI) and latterly \emph{data analytics}, have been identified as major commercial and technological developments that cloud computing can host and enhance. Both of these technologies provide ways of analysing data in a meaningful manner so as to facilitate decision making, and are aimed at increasing productivity and enhancing business performance\cite{alaqrabi2014}. Increased connectivity between information systems across the world have grown globally integrated databases with higher complexities than the traditional organisational repositories.These databases are often spread across cloud computing infrastructure, giving rise to new opportunities in web services-based business intelligence, OLAP using XML data files\cite{alaqrabi2013,alaqrabi2013a,alaqrabi2012}, and predictive analytics as multidimensional data organisations emerge along with advances in technology at lower cost.

Online analytical processing (OLAP) is the user-end interface of BI that is designed to present multi-dimensional graphical reports to the end users. OLAP employs a technique called multidimensional analysis and is mainly used to enable flexible interactive analysis of multidimensional data\cite{chaudhuri2011}. Predictive analytics extends data analytics to incorporate modelling and learning capabilities to facilitate the forecasting of all aspects of an organisation and its business environment.

A BI, OLAP and data/predictive analytics framework is expected to provide timely, accurate, organised and integrated information to business decision makers\cite{glaser2008}.

Cloud BI is the concept of delivering BI and data analytics capabilities as a service. With cloud BI solutions, business users will be able to keep a better fiscal control over IT projects and have the flexibility to elastically scale up or down usage as needs change\cite{alaqrabi2014}.
\subsection{Heterogeneous service delivery and security}
Currently, distributed data analytics applications in business are encompassing an increasing level of computation and a similarly increasing level of enthusiasm as the benefits of predictive modelling are realised. Generally, a business procedure should be flexible in both application and method to enable dynamic business response. An effect of this is that the execution sequence of a given process may not be predictable for all circumstances, even to the extent that sometimes, the real process of execution can be a ``one-of-a-kind" \cite{Georg2005}.

Thus, the services and applications employed in a procedure are characteristically heterogeneous and might be offered and maintained by various, traditionally unrelated organisations.

Companies have their own homogenous security mechanisms and policies to protect their local resources against security threats but applications residing on the resources of different organisations operate in correspondingly numerous different heterogeneous security realms.

A security realm may be viewed as a group of principals (like people, services, and computers) registered with a certain authentication authority (a trusted principal), and controlled through a consistent set of security processes and policies to marshall access to services and resources \cite{xu2012}. An authentication authority is a principal that is considered universally trusted and can be relied upon to execute trustworthy authentication functions\cite{clercq2002}.

As shown by this explanation, authentication is vital for each security realm and before a principal can have a right to use the resources controlled by a security realm, verification of its identity must be confirmed by the authentication procedure of the security realm in order to ascertain the principal who it purports to be.

To identify a principal during the process of authentication, the principal has to announce credentials that were provided to it by the authentication authority of the security realm.

Cloud-based business processes contain collaborating BI/data analytics services from multiple heterogeneous security realms which need to be executed and engaged dynamically at runtime. If authentication relationships are established among different security realms, the process may involve large numbers of extra and computationally expensive steps for converting artefacts.

Hada et al\cite{hada2002} demonstrate the need for a multiparty session authentication protocol, if a multiparty session is constructed out of multiple two-party sessions. It is difficult in some cases for a session user to determine and verify whether the service instance the user contacts is a member of the same multiparty session.
\subsection{Cross-realm authentication}
In a cross-realm authentication scenario, the techniques used in a two-party session do not address such heterogeneous cross-realm authentication issues, which require both credential conversion and the establishment of authentication paths. 
      
Federated authentication establishments may require time consuming activities for negotiations and amendments. This situation may become further complicated when multiple parties coordinate within an authentication system to access resources hosted on multiple clouds.

Existing frameworks do not address the scenario where members of multiple sub-domains want to interact to access resources stored on multiple clouds. This is also a significant issue for any system where it is envisaged that multiparty authentication is required.

The continued increase in the proliferation of small, discrete compute and storage devices means that cross-realm authentication is required beyond the use case of multi-tenant, multi-cloud infrastructures.

The Internet of Things is one domain where it is foreseen that the ability to orchestrate trusted services from myriad connected devices is essential. Further evidence of such thinking is the research community's interest in microservice architectures, where containerisation is utilised to assist the design of software architectures that can scale rapidly\cite{shadija2017}, which is a fundamental design requirement of IoT applications\cite{hill2015}.
\subsection{Contribution and article organisation}
In this paper, we propose a novel multiparty authentication framework for securing business data analytics within cloud computing infrastructure. More specifically, our proposal applies to the situation where members of different security realms need to access distributed analytics services through a trusted principal.

Our proposed framework can authenticate dynamically, minimising the requirement for additional credential conversion processes that will initiate extensive invocations to intermediate services.

The main contribution has been to propose a system and procedure which simplifies the authentication process that is undertaken, between two unrelated secure business environments, prior to allowing data interchange.

This simplification significantly improves the speed of authentication, thereby reducing overall transaction time, and most notably is achieved without compromising the security of either party.

The rest of the paper is organised as follows: Section 2 reviews the existing solutions of the multiparty authentication. This is intended to identify research gaps in this domain. Section 3 describes the design of the multiparty authentication system for BI, the mechanisms for hosting it on cloud computing. Section 4 presents the results of simulations, and critical analysis of results. Section 5 provides the conclusion of the paper and outlines future work.
\section{Multiparty Authentication}
Cloud computing offers shared applications and software platforms and infrastructure services on multi-tenant information technology systems built upon virtualised infrastructure. The such cloud solutions optimising resource sharing while providing isolation solutions at different levels required by the tenant\cite{li2012}. 

However, in the shared domains of cloud computing, BI and data analytics services are exposed to security and privacy threats by virtue of exploits, eavesdropping, distributed attacks, malware attacks, and other known challenges to cloud computing\cite{alaqrabi2014} and distributed systems in general.
\subsection{Dynamic authentication}
In a multi-tenancy environment, authentication frameworks cannot be static. Business processes and systems running on clouds are dynamic and hence the authentication interactions need to be able to support dynamic behaviour as well.

Burrows et al\cite{burrows1990} proposed a global authentication register comprising a privacy framework for tenants. The register holds a private key and personalised data of each tenant for certification of a registration request. The system issues a private key after the personalised data has been provided to the registrar and matched successfully.

In cloud computing, a global authentication register may be viewed as a multi-tenant database holding primary records about registered tenants and additional extended tables for recording unique detail about each tenant\cite{chen2013}.

The primary table comprises details generated by the cloud service provider created for each tenant as per a standard format. Extended tables may record unique details provided by the tenants against a list of metadata classes (for example, names of spouse, dog, first school, mother, favourite movie star, favourite colour, etc.).

The extension tables assist in validating private details about a tenant before issuing a private key for accessing the user's workspace. This concept is referred to as ``identity-based cryptography"\cite{pippal2013}. The root key is the public key for unlocking a cloud or grid-based workspace.

Root keys are appended with the private key based on an identity-based signature generated by an interaction process, that comprises identity information entered by the client and the corresponding digital signature generated by the server. The signature is used by the authentication registry server to append private key fields in the root key and issue to the requesting client.

\begin{figure}[htb]
 \includegraphics[width=\linewidth]{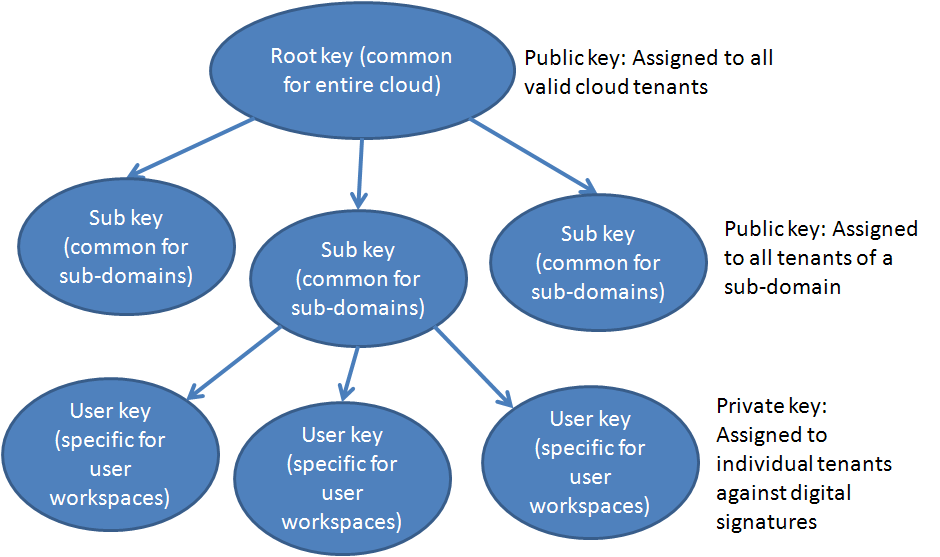}
 \caption{Hierarchical key structures in cloud computing.}
 \label{fig:hierkey}
\end{figure}

In cloud computing, identity-based cryptography and identity-based signature generation may be performed by a separate array of servers (cloud array) dedicated for privacy-as-a-service\cite{pippal2013}. The clouds may comprise hierarchical key structure as shown in Figure \ref{fig:hierkey}.

%


The key has two parts: a public key field and one part as a private key field\cite{chen2013}.

The first public key part is common for all valid cloud tenants whereas the second public key part is common for all tenants of a sub-domain within the main cloud. The private key part is for individual tenants that are determined as per individual digital signatures, which are generated by a separate array of servers against private information provided by the tenants.
\subsection{Cloud sub-domains}
A cloud sub-domain is a group of multiple private virtual workspaces owned by an organisation or a group of related tenants (for example, a community or society). A common public key for the sub-domain ensures that only the tenants owning workspaces are allowed to access them\cite{li2009}.

For example, employees of an organisation will obtain access to the sub-domain public key by virtue of their employment records in the organisation. On top of the sub-domain public key, a tenant-specific private key will be assigned and appended based on private information provided by individual tenants.

Figure\ref{fig:hiermulti} shows how this hierarchical scheme can ensure authentication of employees for uploading or downloading business data to and from the cloud respectively. Private keys for employees may be hosted locally by the company that appends the private key portion with the sub-domain key generated by the cloud for the company. These two keys are finally appended to the root key of the cloud.
\subsection{Cross-cloud federation}
This framework may become complicated further when multiple parties coordinate within an authentication system for accessing resources stored on multiple clouds through the home cloud. In this scenario, a cross-cloud federation system may be established comprising discovery agents, match-making agents and authentication agents\cite{li2011}.

\begin{figure}
  \includegraphics[width=\linewidth]{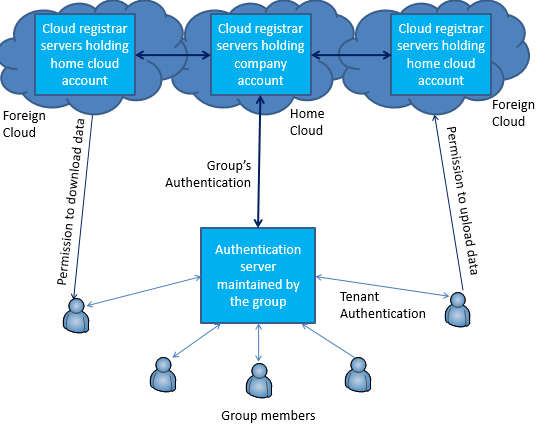}
  \caption{Hierarchical multi-party structure in Multi-cloud computing}
  \label{fig:hiermulti}
\end{figure}

The discovery agent manages a process for discovering the resources requested by the parties by browsing the available foreign clouds. A match-making agent manages a process for short-listing foreign clouds to gain access to the resources requested. The authentication agent creates a security context and trust relationship between the parties and the home cloud, and between the home cloud and the foreign clouds, as shown in Figure \ref{fig:hiermulti}.

There are multiple steps involved in this framework as follows:
\begin{enumerate}
\item All parties make authentication requests.
\item Authentication agent sends queries for artefacts to all parties.
\item Artefact requests are resolved and a group-level public key is generated for all parties.
\item Discovery agent locates resources amongst available foreign clouds.
\item Match-making agent shortlists the preferred foreign clouds.
\item Home cloud requests public keys from the foreign clouds.
\item Foreign clouds make a request for artefacts before sending the keys, which the home cloud provides (establishing trust relationships between the home cloud and the foreign clouds). 
\end{enumerate}

On receipt of the keys, multiple keys are generated integrating the home cloud public key and the corresponding foreign cloud keys for trusted access to the resources.

The resources are collected by the home cloud and made accessible to the authentication server for the entire multi-party group. The individuals in the party gain access to the resources through the authentication server by providing individual secrets and gain private keys. The private keys are appended with the group key that in turn is appended with the home cloud key.

As an alternative to the previous two steps, the home cloud can forward the group's artefacts to the foreign clouds such that they can provide individual public keys to the group's authentication server. These group public keys (assigned by different foreign clouds) can be appended with the individual private keys such that they can be given access to the resources on an individual basis, being a member of this multi-party group.

Such a group authentication protocol requires the sharing of a number of security attributes, like secrets for group session keys, secrets for private keys, secrets for key duplication (for resilience), secrets for session forwarding to other clouds, and secrets for control of key distribution (opening a vault of keys)\cite{li2011}. To facilitate automated inter-organisational processes, trust, dependability and security needs have to be ascertained\cite{qin2013}.

Dependability and security are interrelated through a number of attributes (confidentiality, integrity, availability, reliability, maintainability, and safety). There can be faults and errors in inter-organisational authentication processes leading to failure of services\cite{qin2013}. There are higher chances of faults and errors in modern distributed business environments in which the business specifications are dynamic and the runtime executing them can sometimes be unpredictable. It is very difficult to standardise the runtimes as there may be ``one-of-a-kind", bespoke execution of processes\cite{xu2012}. 

\section{A System for Multiparty Authentication}
\begin{figure}
 \includegraphics[width=\linewidth]{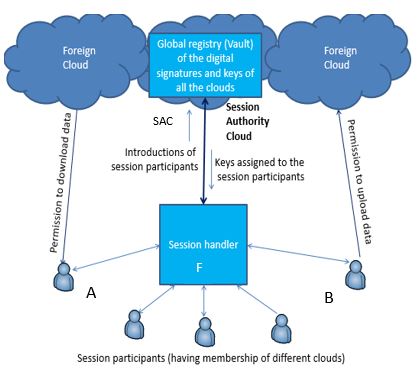}
 \caption{Proposed multiparty session authentication framework in cloud environments.}
 \label{fig:promod}
\end{figure}

%
In this section, the proposed framework for dynamic authentication interactions in a distributed environment is shown in Figure \ref{fig:promod}. We propose the addition of a session authority cloud with the purpose of controlling sessions of multiple clouds.

There shall be no concept of \emph{home} or \emph{foreign} clouds. Every cloud obeys the decisions made by the session authority cloud. The session authority cloud shall hold a large array of servers acting as a security vault.

This vault holds authentication credentials and digital signatures of the tenants of all clouds. The root keys of all the clouds are stored in the vault having folders identifying the clouds. An active tenant will ``know” the root key of its own cloud.

The heterogeneous security realms (sub-domains) are distributed among all of the clouds and are identified through separate sub-domain keys. These keys are stored in subfolders within the corresponding cloud folders. The concept of private key assignment against a digital signature will be adopted, but only for entry into the relevant sub-domain of the cloud.

In the multiparty session scenario, members of multiple sub-domains may interact within a session. All such sessions will be identified by the session authority cloud. The session keys will comprise of a root key of the cloud, sub-domain key, and the portion identifying the session.

This means that there will be multiple session keys valid for a session, each having a common field for the session, but varied fields for cloud root keys and sub-domain keys.

There is no need for any negotiation among the clouds because the session authority cloud ``knows” all of the clouds and their sub-domains. The schematic of the proposed framework is shown in Figure\ref{fig:promod}.
%
%
The fundamental principles of the proposed model are:
\begin{enumerate}[label=(\alph*)]
\item Each session participant should be a tenant of at least one cloud in the multi-cloud framework controlled by the session authority cloud. 
\item If a potential participant is not a cloud member, the introducing participant will have to share credentials with it for joining its own cloud.
\item Each session will have multiple valid keys. While the session key field will be common (refreshed on change of number of participants), the cloud root keys and sub-domain (security realm) keys will vary depending upon the membership profiles of the participants. 
\end{enumerate}
\subsection{A model for secure multiparty authentication}
In this section, we present a multiparty authentication system model for securing BI and data analytics services on the cloud. The proposed system specifically addresses scenarios where business applications that reside on the clouds, that are composed of members of different security realms, want to access distributed data analytics services through a trusted principal. These scenarios are applicable when there are no direct authentication relationships between the stakeholders of different security realms and the distributed data analytics services in multiple cloud systems.
\subsection{Modelling environment and simulation design}
The entire multi-cloud model was created with OPNET modeller, using HP9000 Superdome 64 CPU mega modular servers. These servers are the most powerful systems available in OPNET's model library in the academic edition. Each server can host hundreds of virtual machines. The model is shown in Figure \ref{fig:multimodel} and is described as follows:
\begin{figure*}
  \includegraphics[width=\textwidth]{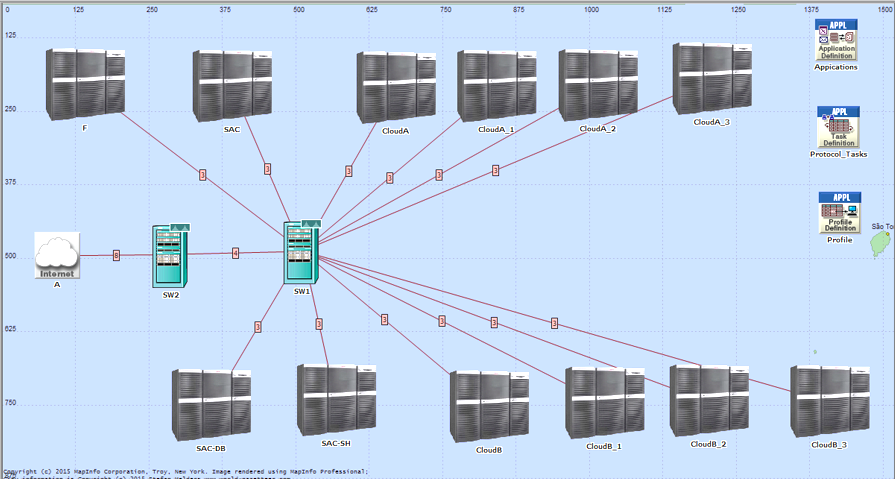}
  \caption{Multi-party authentication model in OPNET.}
  \label{fig:multimodel}
\end{figure*}
\begin{itemize}
\item The object $A$ in this model comprises an Internet cloud of 1000 users, each representing a trusted principal by the $SAC$ requesting a session each for a foreign user (member of a different security realm) on their behalf.
\item The objects $F, SAC, SAC-DB$, and $SAC-SH$ are independent HP9000 Superdome servers, whereas clouds $A$ and $B$ are collections of four HP9000 Superdome servers each.
\item $SW1$ is an Internet switch connecting the trusted principals to the entire inter-cloud framework.
\item $SW2$ is an internal switch of the inter-cloud framework interconnecting the clouds $A$ and $B$ with the core cloud certification authority ($SAC$ in this model) and all its supporting services ($F, SAC-DB$, and $SAC-SH$).
\item $SW1$ and $SW2$ are advanced Cisco chassis-based switches. The red lines represent 1000 BaseX links and the numbers at the centre of each line represent the number of links per connection.
\end{itemize}
For example, the red line connecting $A$ with $F$ has eight 1000 BaseX links. This represents a powerful network within OPNET modeller, with little scope for link and node level congestions for 1000 users connecting the network. It follows that any delays discovered in the simulation results are because of the execution times of the phases of the authentication algorithm.
\subsection{Phase modelling}
The OPNET tasks object has been used to define the phases of the authentication algorithm. Tables \ref{fig:nottimeout} and \ref{fig:timesout} illustrate how they have been modelled within OPNET modeller. Table \ref{fig:nottimeout} illustrates the phases configured without any phase-wise timeouts.

Table \ref{fig:timesout} illustrates a timeout of 60 seconds per phase. Both of the configurations were simulated separately in OPNET modeller. The phases are sequential and each is considered as a request or a response between the stated nodes. In each phase, an appropriate amount of data is transferred as may be needed during practical operation of the phase.

For example, the first phase $A > F$ and the second phase $F > A$ are requesting phases in which the data transmission size is configured as 1024 bytes. However, the third phase $A > F$ is a responding phase ($A$ submits $IDr$ and $IDs$ to $F$) and hence the data transmission size is 4096 bytes.

\begin{table*}[tbp]
\centering
\caption{ALGORITHM STEPS CONFIGURED AS INDIVIDUAL PROTOCOL TASKS IN OPNET TASKS CREATOR OBJECT WITH NO TIMEOUT.}
\label{fig:nottimeout}
\resizebox{\textwidth}{!}{%
\begin{tabular}{@{}llllllll@{}}
\toprule
\textbf{Phase Name}                                                      & \textbf{Start Phase After}   & \textbf{Source} & \textbf{Destination} & \textbf{REQ/RESP Pattern}                & \textbf{End Phase When} & \textbf{Timeout Properties} & \textbf{Transport Connection} \\ \midrule
A\textgreater F:Secure (Request, R1, R2)                         & Application Starts  & A      & F           & REQ\textgreater RESP\textgreater & Final Response & Used           & Default              \\ \midrule
F\textgreater A:Secure (Request, IDr, IDs)                       & Previous Phase Ends & F      & A           & REQ\textgreater RESP\textgreater & Final Response & Used           & Default              \\ \midrule
A\textgreater F:Secure (Response, IDr, IDs)                      & Previous Phase Ends & A      & F           & REQ\textgreater RESP\textgreater & Final Response & Used           & Default              \\ \midrule
F\textgreater SAC: Fetch (R1, R2): IF Valid (IDr, IDs)           & Previous Phase Ends & F      & SAC         & REQ\textgreater RESP\textgreater & Final Response & Used           & Default              \\ \midrule
F\textgreater SAC-DB: Verify (IDr, IDs)                          & Previous Phase Ends & SAC    & SAC-DB      & REQ\textgreater RESP\textgreater & Final Response & Used           & Default              \\ \midrule
SAC-DB\textgreater SAC: Valid (IDr, IDs)                         & Previous Phase Ends & SAC-DB & SAC         & REQ\textgreater RESP\textgreater & Final Response & Used           & Default              \\ \midrule
SAC\textgreater SAC-SH: Invoke (Key, IDess): Fetch (R1, R2)      & Previous Phase Ends & SAC    & SAC-SH      & REQ\textgreater RESP\textgreater & Final Response & Used           & Default              \\ \midrule
SAC:SH\textgreater CloudA: Secure (Access, R1)                  & Previous Phase Ends & SAC-SH & CloudA      & REQ\textgreater RESP\textgreater & Final Response & Used           & Default              \\ \midrule
CloudA\textgreater SAC : SH : Secure (Access, R1)                & Previous Phase Ends & CloudA & SAC-SH      & REQ\textgreater RESP\textgreater & Final Response & Used           & Default              \\ \midrule
SAC:SH\textgreater CloudB: Secure (Request, R2): IF Key (IDsess) & Previous Phase Ends & SAC-SH & CloudB      & REQ\textgreater RESP\textgreater & Final Response & Used           & Default              \\ \midrule
CloudB\textgreater SAC: SH: Secure (Access, R2)                  & Previous Phase Ends & CloudB & SAC-SH      & REQ\textgreater RESP\textgreater & Final Response & Used           & Default              \\ \midrule
SAC: SH\textgreater F: Secure (Access, R1, R2): Key (IDsess)     & Previous Phase Ends & SAC-SH & F           & REQ\textgreater RESP\textgreater & Final Response & Used           & Default              \\ \midrule
F\textgreater A: Secure (Access, R1, R2): Key (IDsess)     & Previous Phase Ends & F & A           & REQ\textgreater RESP\textgreater & Final Response & Used           & Default              \\ \bottomrule
\end{tabular}%
}
\end{table*}
%
The data sizes have been configured accordingly in all phases of the algorithm. A subsequent phase does not begin unless the previous phase has ended. Thus, if a phase fails to complete, the subsequent phase will not begin at all.

A phase will be deemed as completed only when a final response has arrived from the requested node to the requesting node. 
\begin{table*}[tbp]
\centering
\caption{TIMEOUT INTRODUCED IN EACH PHASE OF THE AUTHENTICATION PROTOCOL.}
\label{fig:timesout}
\resizebox{\textwidth}{!}{%
\begin{tabular}{@{}llllllll@{}}
\toprule
\textbf{Phase Name}                                                      & \textbf{Start Phase After}   & \textbf{Source} & \textbf{Destination} & \textbf{REQ/RESP Pattern}                & \textbf{End Phase When} & \textbf{Timeout Properties} & \textbf{Transport Connection} \\ \midrule
A\textgreater F:Secure (Request, R1, R2)                         & Application Starts  & A      & F           & REQ\textgreater RESP\textgreater & Final Response & Not Used           & Default              \\ \midrule
F\textgreater A:Secure (Request, IDr, IDs)                       & Previous Phase Ends & F      & A           & REQ\textgreater RESP\textgreater & Final Response & Not Used           & Default              \\ \midrule
A\textgreater F:Secure (Response, IDr, IDs)                      & Previous Phase Ends & A      & F           & REQ\textgreater RESP\textgreater & Final Response & Not Used           & Default              \\ \midrule
F\textgreater SAC: Fetch (R1, R2): IF Valid (IDr, IDs)           & Previous Phase Ends & F      & SAC         & REQ\textgreater RESP\textgreater & Final Response & Not Used           & Default              \\ \midrule
F\textgreater SAC-DB: Verify (IDr, IDs)                          & Previous Phase Ends & SAC    & SAC-DB      & REQ\textgreater RESP\textgreater & Final Response & Not Used           & Default              \\ \midrule
SAC-DB\textgreater SAC: Valid (IDr, IDs)                         & Previous Phase Ends & SAC-DB & SAC         & REQ\textgreater RESP\textgreater & Final Response & Not Used           & Default              \\ \midrule
SAC\textgreater SAC-SH: Invoke (Key, IDess): Fetch (R1, R2)      & Previous Phase Ends & SAC    & SAC-SH      & REQ\textgreater RESP\textgreater & Final Response & Not Used           & Default              \\ \midrule
SAC:SH\textgreater CloudA: Secure (Access, R1)                  & Previous Phase Ends & SAC-SH & CloudA      & REQ\textgreater RESP\textgreater & Final Response & Not Used           & Default              \\ \midrule
CloudA\textgreater SAC : SH : Secure (Access, R1)                & Previous Phase Ends & CloudA & SAC-SH      & REQ\textgreater RESP\textgreater & Final Response & Not Used           & Default              \\ \midrule
SAC:SH\textgreater CloudB: Secure (Request, R2): IF Key (IDsess) & Previous Phase Ends & SAC-SH & CloudB      & REQ\textgreater RESP\textgreater & Final Response & Not Used           & Default              \\ \midrule
CloudB\textgreater SAC: SH: Secure (Access, R2)                  & Previous Phase Ends & CloudB & SAC-SH      & REQ\textgreater RESP\textgreater & Final Response & Not Used           & Default              \\ \midrule
SAC: SH\textgreater F: Secure (Access, R1, R2): Key (IDsess)     & Previous Phase Ends & SAC-SH & F           & REQ\textgreater RESP\textgreater & Final Response & Not Used           & Default              \\ \bottomrule
\end{tabular}%
}
\end{table*}
%
Thus, if there is congestion on the network and a timeout has been configured for each phase, the session will be dropped if any of the subsequent phases fails to execute successfully.
\subsection{Simulation parameters and assumptions}

The remaining assumptions for the simulation are as follows:
\begin{enumerate}[label=(\alph*)]
\item The phases have been packaged in an application called as ``Protocol\_Tasks" in the model. Protocol\_Tasks is a custom application, which in turn is designed using the tasks object in OPNET modeller. One needs to enter the attributes in this field and select the task object packaged with all the phases configured.
\item The database application has been configured for running on $SAC-DB$ only. OPNET's default high load task format has been employed. There was no need to configure it manually because it is not the focus of this research.
\item The applications are executed using the profiles object, as shown in the following table. Protocol\_tasks and $SAC-DB$ have been configured to execute independently of each other such that they do not cause a conflict during simulation. Both applications have been assigned a start offset of 5 to 10 seconds.
\end{enumerate}
However, the start offset of the network itself has been configured at 100 to 110 seconds. This is because the network is large and should be given enough time to complete the tasks of the routing protocol. In this model, the routing protocol selected is RIPv2 (which is OPNET's default).

Before explaining the simulations, it is essential to clarify how the nodes recognise whom to contact to execute a phase. This clarification is needed to map the symbols ($A, F, SAC, SAC-DB$, Cloud $A$, and Cloud $B$) with actual servers (names of servers in the network) on the network. For simplicity, the symbols and server names have been kept the same in the model.

However, in reality, the cloud environment will be completely different. This is a complex design in OPNET that needs to be configured carefully by matching the source-destination relationships with the phases of the customer application (Protocol\_Tasks authentication algorithm).

In destination preferences, a node is allowed to communicate with only those nodes that it is supposed to interact with for executing the algorithm. Hence, $A$ is allowed to communicate only with $F$, and $F$ is allowed to communicate only with $A$ and $SAC$.

It should also be noted that a recursive relationship ($A$ communicating with $A$ and $F$ communicating with $F$) is established in the destination preferences as well. This ensures that the node is able to communicate with itself whenever required by OPNET.

Finally, the attachment of a profile to a node is important to provide instruction about what it needs to execute. This is done by identifying the profile within the node configuration. In this model, all nodes are configured to execute the ``Protocol\_Tasks" whereas $SAC-DB$ is configured to execute ``Protocol\_Tasks" as well as the database application.
\subsection{Authentication approval protocol}
The proposed protocol is specifically addressed to scenarios of BI and data analytics applications that are accessed via clouds, where members of different security realms want to access distributed analytics services through a trusted principal.

These scenarios are applicable when there are no direct authentication relationships between the people of different security realms and the distributed BI services in multiple cloud systems.

The session approval protocol begins with a user having membership in any security realm that the trusted principal recognises. It is assumed to provide access to the BI database objects in clouds $A$ and $B$, if the $SAC$ approves the request forwarded by the principal.

It is also assumed that $SAC$ will not entertain any request that is not forwarded by the principal. The user requesting access is neither a member of $Cloud A$ nor a member of $Cloud B$. In essence, the user is a member of a security realm that is a different cloud ($Cloud C$), which is trusted by the $SAC$ (which means that the third cloud is a member of the $SAC-DB$). Most importantly, the principal should recognise who is the user because the $SAC$ trusts the principal for accepting the session request.

The steps in the algorithm for the authentication protocol are shown in Table\ref{fig:nottimeout}. $IDr$ is the cloud membership key of the requesting user. $IDs$ is the sub-domain membership of the requesting user, and $IDsess$ is the session key assigned by the $SAC$ for accessing the BI database files residing on clouds $A$ and $B$.

Clouds $A$ and $B$ will open the access to this key only as it is approved and forwarded by the $SAC$ (through $SAC-SH$). $R1$ and $R2$ are BI database files residing on clouds $A$ and $B$ that are requested by the requesting user. $A$ is the trusted principal through which, the requesting user has approached the $SAC$. 
\section{Analysis of Results}
\begin{figure}[h]
 \includegraphics[width=\linewidth]{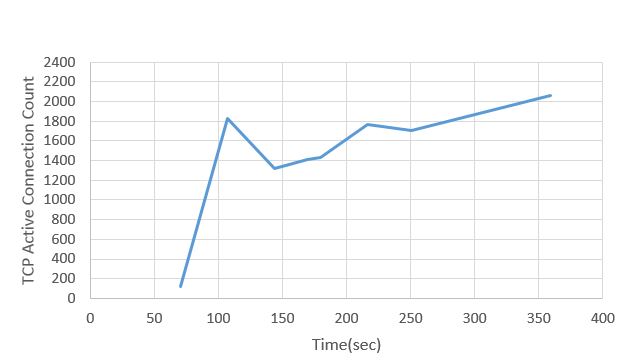}
 \caption{TCP sessions initiated by node ``A"}
 \label{fig:tcpsess}
\end{figure}

\begin{figure}
  \includegraphics[width=\linewidth]{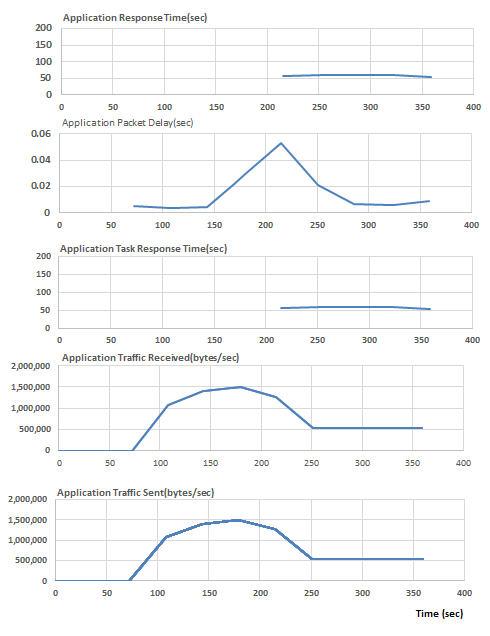}
  \caption{Overall performance metrics and behaviours of the authentication protocol tasks on the network.}
  \label{fig:overperf}
\end{figure}

We observed that the simulation executed 178 million events in the operation of seven minutes and nine seconds on the network. This is because it is a reasonably large network with 1000 trusted principals accessing; a near real-world scenario and hence the results have practical significance. Figure \ref{fig:tcpsess} illustrates the TCP sessions initiated by the node ``$A$". Active TCP sessions exceeded 2000 during the simulation period, indicating that each trusted principal has made two session requests on average. Hence the authentication protocol has been triggered more than 2000 times on the network.
%
The overall performance and behaviour of authentication protocol tasks on the network are shown in Figure \ref{fig:overperf}.

The first statistic shows that the overall (end-to-end) response time of the authentication protocol on the network is about 60 seconds. This is genuine given the time taken in establishing the TCP connection and transferring the data.

Hence, 60 seconds is a committed performance for executing all of the 13 phases. However, this is feasible knowing that it is only a one-time activity for each user being added by the trusted principal. 
%
%
It is important to note that delays caused by network or host-based congestions are more serious than protocol execution delays. This has been verified by confirming the packet network delay measurement results.

It is observed that the maximum delay occurrence on the network is less than 0.06 seconds. It was further confirmed by device specific reports that none of the servers, switches, and links had registered any packet queues or packet forwarding delays on the network. Hence, 60 seconds is a committed performance for executing all of the 13 phases. This delay has occurred because of the amount of data exchanged per phase. The metric ``application task response time" is the same as ``application phase response time" in this model, because each phase has only one task in the algorithm. The last two metrics reflect the overall authentication traffic sent and received on the network. So, for 1000 trusted principals interacting, the traffic of up to 1.5 Mbps is quite moderate. This shows that the protocol is loading the network moderately.

\begin{figure}
  \includegraphics[width=\linewidth]{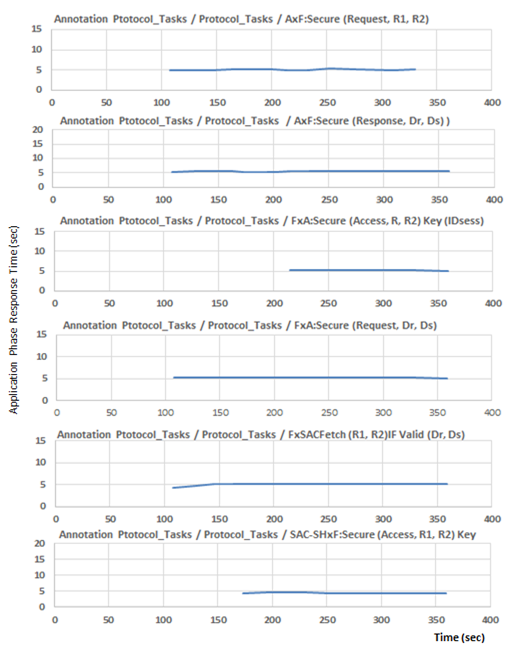}
  \caption{Response times of the individual phases of the authentication protocol.}
  \label{fig:resptim}
\end{figure}

Before drawing any final conclusions, the response times of individual phases of the authentication protocol were also investigated. Figure \ref{fig:resptim} illustrates this metric for six out of the eleven phases of the protocol. The overall response time is 5 seconds per phase. This is genuine given the time taken in establishing the TCP connection and then transferring the data. The above results pertain to the phases executed without a timeout.
%
A second simulation was carried out, including a timeout of 60 seconds per phase. As per the previous results, the phases should not time out given that each phase was taking about 5 seconds to execute.However, as shown in Figure \ref{fig:overperf}, the number of application instances reduced significantly as the phases of the protocol progressed. This indicates massive session drops because of timeout configuration.

This result was unexpected because the timeout configured per phase was larger than the average phase duration observed. Multiple settings were tested but the results were similar and is a problem needing further investigation.

In this research, it is recommended that the 13-step authentication protocol should not have any timeout configured. However, a localised timeout can be configured at $F$. If the responses from the cloud are not received within 200 seconds, the session may be dropped automatically. This means the full authentication process is from 0-200 seconds. Clouds are like galaxies of servers. For example, users from multiple companies sitting on multiple clouds may collaborate on a common research project with the project manager acting as the trusted principal. The SAC can guarantee secured access to resources on different clouds. However, it cannot guarantee performance and committed response times, given that each member cloud may have its own network configurations. While such an authentication protocol is essential for multiparty collaborations on multiple clouds, timeout settings may not be feasible for all the phases of the authentication protocol. 
\section{Conclusions and Future Work}
The cloud computing paradigm is a fundamental, enabling model for the future development of BI and data analytics services. Cloud platforms offer several distinct advantages in terms of cost efficiency, reliability, flexibility and scalability of implementation.

Moreover, the inherent feature of connectivity of cloud architecture offers the opportunity to take advantage of enhanced data sharing capabilities. In this article, we have proposed a novel authentication mechanism model for multi-party authentication of BI mechanisms for hosting on cloud computing.

This article has considered problems associated with reliable, timely and secure data transfer mechanisms necessary for shared business data processing networks.
This multiparty authentication system for dynamic authentication interactions is effective when members of different security realms want to access distributed business data services through a trusted principal.

Our proposed mechanism can help cloud session users authenticate their session membership so as to largely simplify the authentication processes within multi-party sessions. While this paper has presented the framework, additional research and development is needed to develop a set of protocols for multi-party session management and cross-realm authentication for dynamic authentication interactions between users and data services in multiple cloud systems located in different security realms.

The scenario of multiparty authentication across security realms is not limited to business processes that require access to data services. Internet of Things (IoT) architectures are a contemporary example of a need to be able to model, comprehend and deploy authentication mechanisms that can securely tolerate myriad network nodes that each provide more cohesive services\cite{hill2015}.

We are now progressing the work in two ways. First, we are formally evaluating the authentication protocols in order to robustly deploy the framework to an environment that is composed of a mixed set of discrete devices, enabling the framework to be tolerant of existing and emerging technologies for cloud platforms and network infrastructures. Second, we are now including a variety of devices in addition to clouds within the simulation, to better understand the effects upon network performance when greater numbers of low powered compute and storage wireless nodes are introduced. This replicates the emerging IoT and Industrial IoT scenarios where extensive integration of Wireless Sensor Networks and edge computing nodes is commonplace.

\end{document}